\documentclass[prx,twocolumn,10pt]{revtex4-1} 
\usepackage{amsfonts,amsmath,amssymb,graphicx}

\usepackage{color,times}

\definecolor{darkblue}{rgb}{0, 0, 0.8}
\usepackage[colorlinks=true, breaklinks=true, linkcolor=red, citecolor=darkblue, urlcolor=darkblue]{hyperref}

\newcommand{\ket}[1]{\left|#1\right\rangle}
\newcommand{\bra}[1]{\left\langle#1\right|}

\newcommand{\ave}[1]{\left\langle #1\right\rangle}

\newcommand{\bs}{\boldsymbol}
\newcommand{\dd}{{\rm d}}

\newcommand{\figref}[1]{Fig.~\ref{#1}}
\renewcommand{\eqref}[1]{Eq.~(\ref{#1})}

\begin{document}

\title{Observing the space- and time-dependent growth of correlations \\
in dynamically tuned synthetic Ising antiferromagnets}

\author{Vincent Lienhard$^*$}
\author{Sylvain de L\'es\'eleuc$^*$}
\author{Daniel Barredo}
\author{Thierry Lahaye}
\author{Antoine Browaeys}
\affiliation{Laboratoire Charles Fabry, Institut d'Optique Graduate School, CNRS, \\ 
Universit\'e Paris-Saclay, F-91127 Palaiseau Cedex, France}
\author{Michael Schuler$^*$}
\author{Louis-Paul Henry}
\author{Andreas M. L\"auchli}
\affiliation{Institut f\"ur Theoretische Physik, Universit\"at Innsbruck, A-6020 Innsbruck, Austria}

\date{\today}

\begin{abstract}
We explore the dynamics of artificial one- and two-dimensional Ising-like quantum antiferromagnets with different lattice geometries by using a Rydberg quantum simulator of up to 36 spins in which we dynamically tune the parameters of the Hamiltonian. We observe a region in parameter space with antiferromagnetic (AF) ordering, albeit with only finite-range correlations. We study systematically the influence of the ramp speeds on the correlations and their growth in time. We observe a delay in their build-up associated to the finite speed of  propagation of correlations in a system with short-range interactions. We obtain a good agreement between experimental  data and numerical simulations taking into account experimental imperfections measured at the single particle level. Finally, we develop an analytical model,  based on a short-time expansion of the evolution operator, which captures  the observed spatial structure of the correlations, and their build-up in time. 
\end{abstract}

\maketitle

\begin{figure*}[t!]
\includegraphics[width=16.8cm]{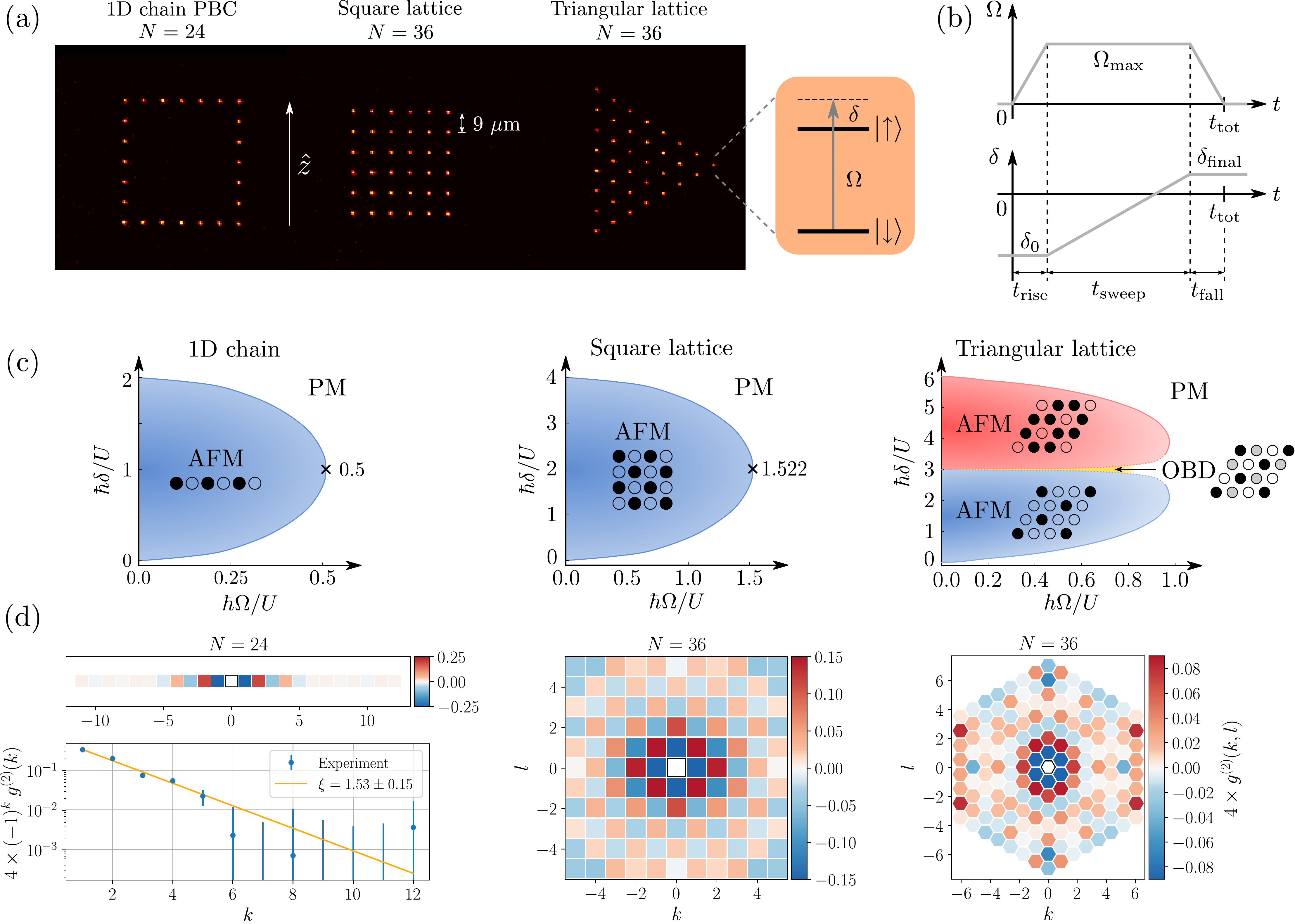}
\caption{Studying the AF Ising model on 1d and 2d systems. (a) Examples of single-shot fluorescence images of single-atom arrays used in our experiments: a 24-atom 1d chain with periodic boundary conditions, a $6\times6$ square array, and a 36-atom triangular array.  Each atom is used to encode a spin-1/2, whose internal states $\ket{\uparrow}$ and $\ket{\downarrow}$ are coupled with Rabi frequency $\Omega$ and detuning $\delta$.  (b) Time dependence of the Rabi frequency $\Omega(t)$ and detuning $\delta(t)$ used to probe the build-up of correlations.  (c) Ground state phase diagrams of the Ising model \eqref{eq:hising},  in the nearest-neighbor interaction limit, for a 1d chain, a 2d square lattice, and a 2d triangular lattice. AFM: antiferromagnetic. PM: paramagnetic. OBD: ``order by disorder". (d) Typical experimental correlation functions obtained for these geometries (see text).  For the 1d chain the correlation length $\xi=1.5$~sites (bottom left panel). }
\label{fig:fig1}
\end{figure*}

\section{Introduction}

The study of non-equilibrium dynamics is currently one of the most challenging areas of quantum many-body physics. In contrast to the equilibrium case, where statistical
physics provides a general theoretical framework and where very powerful numerical methods are available, the out-of-equilibrium behavior of quantum matter presents a
wide variety of phenomena and is extremely hard to simulate numerically, especially in dimensions $d>1$. When the parameters of a quantum many-body system are quenched
abruptly, or more generally ramped with a finite rate into a new quantum phase, correlations and entanglement build up and propagate over the system, which may, at long
times, either thermalize or retain memory of its initial state when many-body localization occurs~\cite{Nandkishore2015,Altman2015}. The speed at which the correlations propagate depends on the range of the interaction and is limited by the Lieb-Robinson bounds~\cite{Lieb1972,Bravyi2006,Eisert2013,Foss-Feig2015,Hazzard2017,SanchezPalencia2016}.

An attractive way to study this physics has emerged in the last years, and consists in using quantum simulators, \emph{i.e.}~well-controlled, artificial quantum systems that implement experimentally the Hamiltonian of interest~\cite{Georgescu2014}. Spin Hamiltonians that are used in condensed-matter physics to describe {e.g.}~quantum magnets are arguably the simplest quantum many-body systems that can be used to study non-equilibrium dynamics: even though they involve distinguishable particles with only internal degrees of freedom, the interplay between interactions, geometry and dimensionality provides a wealth of distinct quantum phases into which the system can be driven. In recent years, many experimental platforms for the quantum simulation of spin Hamiltonians have been developed using the tools of atomic physics. For example, equilibrium properties of synthetic quantum magnets have been studied using trapped ions~\cite{Kim2010,Richerme2013} or ultra-cold atoms in optical lattices~\cite{Simon2011,Sengstock2011,Landig2016,Cheuk2016,Brown2017} including e.g. the  observation of long-range antiferromagnetic order~\cite{Mazurenko2017}. Many experiments using these platforms were also devoted to the study of non-equilibrium dynamics,  including the investigation of the Lieb-Robinson bound~\cite{Cheneau2012,dePaz2013,Richerme2014,Hild2014,Meinert2014,Jurcevic2014,Zhang2017}.

Recently, a new platform, using arrays of individually resolved atoms excited to Rydberg states, has been shown to provide a versatile way to engineer synthetic quantum Ising magnets~\cite{Browaeys2016}. Pioneering experiments in quantum gas microscopes studied quenches \cite{Schauss2012} or slow sweeps \cite{Schauss2015} in a regime where the blockade radius $R_{\rm b}$, i.e. the distance over which interatomic interactions prevent the excitation of two atoms, was much larger than the lattice spacing $a$, rendering the underlying lattice hardly relevant. In this case, the observed correlations are liquid-like, and observing the crystal-like ground state of the system~\cite{Pohl2010} would require exponentially long ramps~\cite{Petrosyan2016}. More recently, experiments with arrays of optical tweezers allowed exploring the regime $R_{\rm b} \gtrsim a$, studying non-equilibrium dynamics following quenches~\cite{Labuhn2016} or slow sweeps~\cite{Bernien2017}.

Here, we use a Rydberg-based platform emulating an Ising antiferromagnet to study the growth of correlations during ramps of the experimental parameters in 1d and 2d arrays of up to 36 single atoms with different geometries.  We operate in the regime $R_{\rm b} \simeq a$ where the interactions act  to a good approximation only between nearest neighbors.
We dynamically tune the parameters of the Hamiltonian and observe the build-up of antiferromagnetic order.  We also observe the influence of the finite ramp speed on the extent of the correlations, and follow the development in space and time of these correlations during a ramp. Numerical simulations of the dynamics of the system without any adjustable parameters are in very good agreement with the experimental data, and show that single-particle dephasing arising from technical imperfections currently limits the range of the observed correlations. Finally, we observe a characteristic spatial structure in the correlations, which can be understood qualitatively by a short-time expansion of the evolution operator for both, square and triangular lattices.

The paper is organized as follows. In Sec.~\ref{Sec:exp_platform}, we describe the experimental platform. After recalling the phase diagram for the different array geometries  (Sec.~\ref{Sec:phase_diag}), we explore it for the square array (Sec.~\ref{Sec:square_phase_diag}). In Sec.~\ref{Ssec:varying_speed} we study the influence of ramp speeds on the correlations. In Sec.~\ref{Ssec:time_dependence}, we observe a delay in the build-up of the spin-spin correlations, a feature linked to their finite speed of propagation. 
Finally, in Sec.~\ref{Ssec:space_dependence}, we analyze the 2d spatial structure of the AF correlations on the square and triangular geometries and show that it is qualitatively captured by an analytical model based on short-time expansion.

\section{Experimental Platform}\label{Sec:exp_platform}

Our experimental platform (see Appendix~\ref{app:exp}) is based on user-defined two-dimensional arrays of optical tweezers, each containing a single $^{87}{\rm Rb}$ atom \cite{Labuhn2016}. Here we use the arrays shown in Fig.~\ref{fig:fig1}(a) containing up to $N=36$ atoms: a 1d chain with periodic boundary conditions (PBC), a square lattice, and a triangular lattice. We achieve full loading of the arrays using our atom-by-atom assembler~\cite{Barredo2016}. The atoms are prepared in the ground state $\ket{\downarrow}=\ket{5S_{1/2},F=2,m_F=2}$ by optical pumping, and then coupled coherently to the Rydberg state $\ket{\uparrow}=\ket{64D_{3/2},m_j=3/2}$ with a two-photon transition of Rabi frequency $\Omega$ and a detuning $\delta$, while the traps are switched off. The system is described by the Hamiltonian:
\begin{equation}
H=\sum_i\left(\frac{\hbar\Omega(t)}{2}\sigma_i^x-\hbar\delta(t)n_i\right)+\frac{1}{2}\sum_{i \neq j}U_{ij}n_in_j,
\label{eq:hising}
\end{equation}
where $n_i=\ket{\uparrow}\bra{\uparrow}_i$ is the projector on the Rydberg state for atom $i$, and $\sigma^x=\ket{\uparrow}\bra{\downarrow}+\ket{\downarrow}\bra{\uparrow}$ is the $x$-Pauli matrix. The interaction term $U_{ij}$ arises from van der Waals interactions between the atoms, and thus scales as $1/r_{ij}^6$ with the distance $r_{ij}$ between atoms $i$ and $j$. This short-range character allows us to neglect interactions beyond nearest-neighbor (NN) atoms for this work~\cite{footnote_nn_only}, and we thus restrict~\eqref{eq:hising} to NN terms only. For the $D$ states we use, the van der Waals interaction is anisotropic \cite{Browaeys2016,deLeseleuc2017b}, and the lattice spacings in the arrays are tuned such that the NN interactions anisotropy is below 10\%. We use typical values $\Omega_\text{max}/(2 \pi) \sim 2 $~MHz and $U/h \sim 1 - 3 $~MHz, see Table~\ref{tab:tab1}. The driving parameters $\Omega$ and $\delta$ can be considered being constant over the entire array (see Appendix~\ref{app:exp}).

The system thus realizes an Ising-like model with a transverse field $\propto\Omega$ and a longitudinal field $\propto\delta$, which gives rise to AF order for $U>0$ (see below). We probe the system by using time-dependent ramps $\Omega(t)$ and $\delta(t)$ as shown in Fig.~\ref{fig:fig1}(b). The Rabi frequency $\Omega$ is switched on and off in $t_\text{rise},\,t_\text{fall} \sim 250$~ns at a constant detuning $\delta$, and, in between, $\delta$ is ramped linearly from $\delta_0/(2 \pi) = -6$~MHz to $\delta_{\rm final}$ during the time $t_\text{sweep}$. The total duration of the ramp is then $t_{\rm tot}=t_\text{rise}+t_\text{sweep}+t_\text{fall}$. After this, the trap array is switched on again. Atoms in $\ket{\downarrow}$ are observed by fluorescence, while those in $\ket{\uparrow}$ are lost from their trap. For a given set of parameters, the experiment is repeated a few hundred times to reconstruct quantities of interest such as the Rydberg density (equivalent to the magnetization), spin-spin correlation functions, or the sublattice density histogram. When $\delta_\text{final}$ lies in the AF region (see Fig.~\ref{fig:fig1}c), correlation functions show the emergence of short-range order [Fig.~\ref{fig:fig1}(d)]  (see more details on the measurements in Sec~\ref{Sec:square_phase_diag}). 

\section{Theoretical Phase Diagrams and State Preparation Considerations}\label{Sec:phase_diag}

The calculated ground state phase diagrams for the nearest-neighbor Ising model are shown in Fig.~\ref{fig:fig1}(c) for the three geometries considered in this paper. On the 1d chain, the phase diagram is well-known and features an AF phase and a paramagnetic phase delimited by a second-order quantum phase transition line of the (1+1)d Ising universality class~\cite{SachdevQuantumPhaseTransitions}. For $\left(\hbar\delta/U\right)_\text{TFI} = z/2$, with the coordination number $z=2$, the Hamiltonian corresponds exactly to the analytically solvable transverse field Ising (TFI) model without longitudinal field, where in 1d the critical point $\left(\hbar\Omega/U\right)_c= 1/2$ is known analytically.

The phase diagram for the NN Hamiltonian on the square lattice ($z=4$) is qualitatively similar to the phase diagram on the chain, with again an AF phase and a paramagnetic phase delimited by a second-order quantum phase transition line in the (2+1)d Ising universality class. For the transverse field Ising line $\left(\hbar\delta/U\right)_\text{TFI}$, the critical point is known to high precision from Monte-Carlo simulations, $\left(\hbar\Omega/U\right)_c=  1.52219(1)$~\cite{Bloete2002}. Finally, on the triangular lattice ($z=6$) the NN Hamiltonian features a much richer phase diagram. A Rydberg crystal with filling fraction $1/3$ (at $\Omega=0$), where one of the three triangular sublattices is occupied by Rydberg states and the atoms in the other sublattices remain in the ground state, appears in a region within $0 < \hbar\delta/U < z/2$. The conjugate crystals obtained by flipping all the spins (filling fraction $2/3$ at $\Omega=0$) lies in the region within $z/2 < \hbar\delta/U < z $. The most interesting feature occurs when $\hbar\delta/U=z/2$, where an ``order by disorder'' process occurs (see Appendix~\ref{app:orderbydisorder}). When including the open boundary conditions for the square and triangular $N=36$ arrays studied here, the ``phase diagrams" present the same phases with some modifications (see Appendix~\ref{app:openclusters}).

In the experiments reported here, we initialize the system in the product state with all atoms in their ground state $\ket{\downarrow}$. Since the system is not at equilibrium with a thermal bath, we cannot simply cool to the ground state. Therefore we use an experimental protocol involving sweeps of $\Omega(t)$ and $\delta(t)$ in order to characterize the ground states. Ideally, we would use adiabatic state preparation to reach the targeted ground state. In order for this approach to succeed, the duration of the sweep, which scales as the inverse of the square  of the energy gap $\Delta$ above the instantaneous ground states~\cite{Das2008},  should be smaller than the coherence time of the system.  In the 1d chain and the 2d square lattice, the minimal gap at the quantum phase transition   scales as the inverse of the linear size of the system: $\Delta \sim 1/N$ in 1d and $\Delta \sim 1/L=1/\sqrt{N}$ in 2d~\cite{SachdevQuantumPhaseTransitions,Schuler2016}. The gap also depends on the excitation velocity at the quantum critical point, which can vary substantially along the quantum phase transition lines.  For the triangular lattice, we expect a first-order quantum phase transition between the paramagnet to either the 1/3 or 2/3 filling Rydberg crystals, resulting in a minimal gap exponentially small in $N$~\cite{Laumann2015}. Due to these scalings, the gaps are small for the number of atoms used here. As a consequence adiabatic state preparation would require long pulses. In the absence of any imperfections such as dephasing, pulse durations $t_{\rm tot}$ of a few $\mu$s would allow to reach strong correlations extending over the entire system. While such durations are experimentally accessible,  state-of-the-art platforms~\cite{Schauss2015,Labuhn2016,Bernien2017} show significant dephasing over these timescales. For this reason, we approach here the question of state preparation from the opposite side: we ask how much correlations and what kind of structures we can expect being built up in a given amount of time. Answering these questions also informs us about the minimal time required to build up highly correlated states starting from a product state~\cite{Bravyi2006}, {\it i.e.}~about a quantum speed limit in our many-body system.

\section{Exploring the Square Lattice Phase Diagram}\label{Sec:square_phase_diag}

\begin{figure*}[t]
\centering
\includegraphics[width=\linewidth]{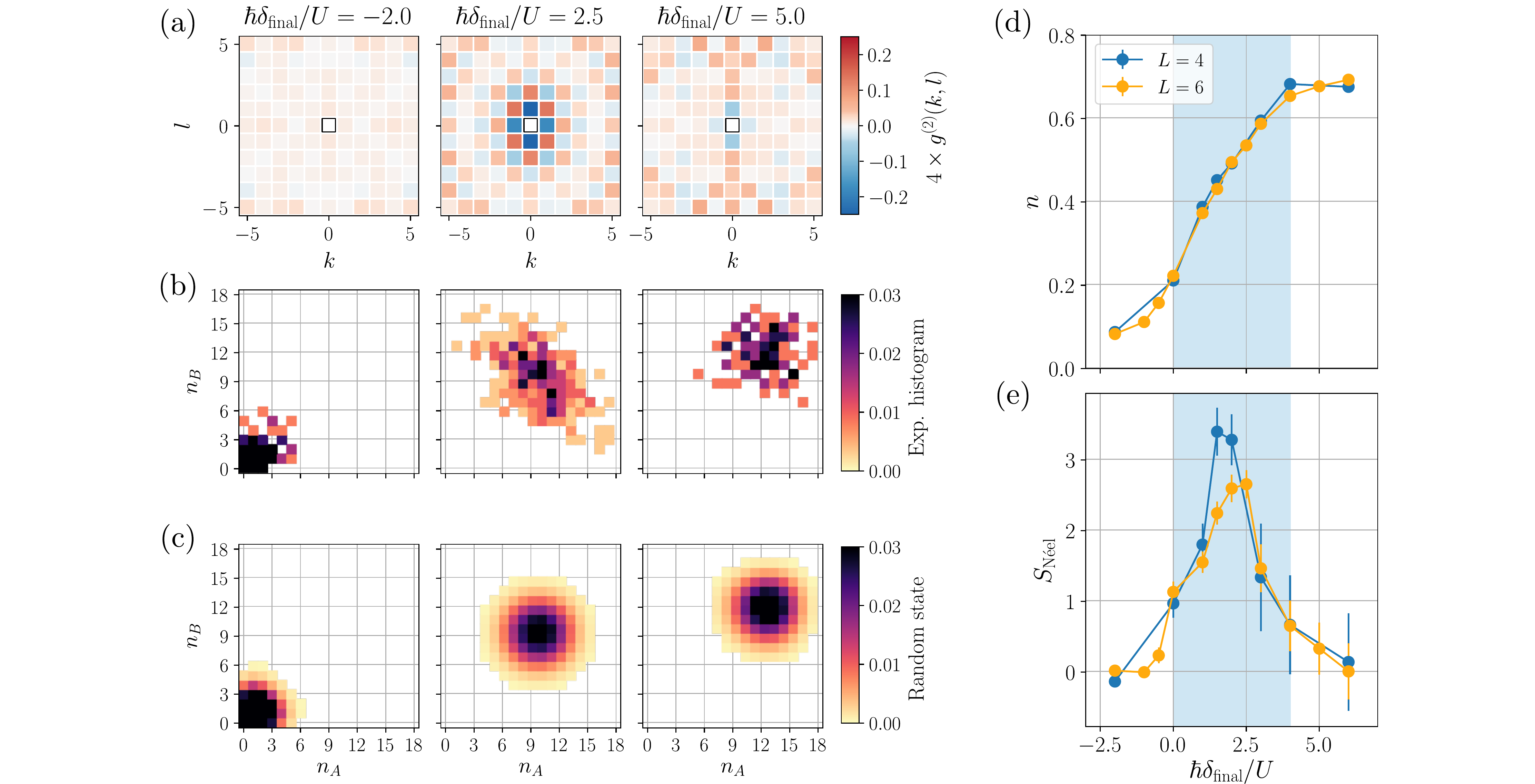}
\caption{Exploring the $\Omega=0$ line of the phase diagram for a $L\times L$ square lattice with $L=4$ and $L=6$. (a) Experimental spin-spin correlation function $g^{(2)}(k,l)$ for different values of the final detuning. (b) Corresponding experimental histograms of the populations of the two N\'eel sublattices $A$ and $B$, to be contrasted with those in (c), calculated for uncorrelated random states with the same Rydberg density. (d) Average Rydberg density $n$, and (e) N\'eel structure factor $S_\text{N\'eel}$ as a function of the final detuning. The shaded areas highlight the region of the AF phase in the phase diagram for $\Omega=0$.}
\label{fig:fig2}
\end{figure*}

\begin{figure*}[t]
\includegraphics[width=\linewidth]{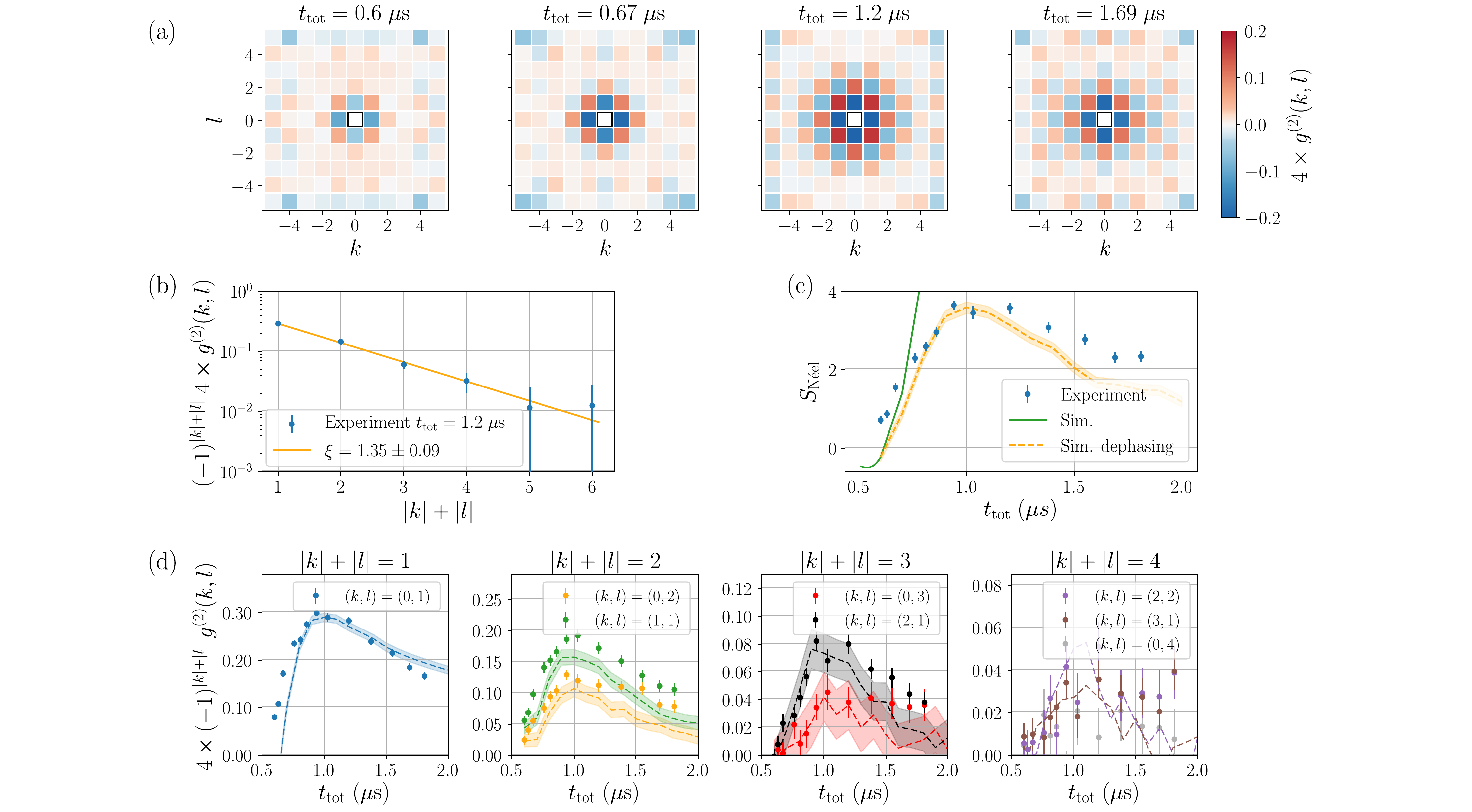}
\caption{Build-up and decay of correlations in a 2d square lattice antiferromagnet for different ramp durations $t_{\rm tot}$. (a) Experimental results for the 2d correlation function $g^{(2)}(k,l)$ for four different values of $t_{\rm tot}$, showing short-range AF ordering, with a pronounced sweep duration dependence. (b) The correlation function $g^{(2)}$ (averaged within a Manhattan shell $m = |k|+|l|$) decays exponentially with the Manhattan distance $m$, with a correlation length $\xi$ of about 1.4 lattice sites. (c) N\'eel structure factor $S_\text{N\'eel}$ as a function of $t_{\rm tot}$. The green solid line shows the result of a numerical simulation using exact diagonalization for a $4\times4$ system without dephasing, while the yellow dashed line denotes a numerical simulation including dephasing (see Appendix~\ref{app:numerics}). (d) Dependence of the correlations on $t_{\rm tot}$ for first-, second-, third- and fourth-neighbors. One observes a dependence not only on the Manhattan distance $|k|+|l|$, but also on $k-l$ (see Sec.~\ref{Ssec:space_dependence}). The dashed lines is a numerical simulation including dephasing (the shaded area corresponds to the s.e.m. obtained therein).} 
\label{fig:fig3}
\end{figure*}

In a first set of experiments we map out the $\Omega=0$ ``phase diagram'' on a $L\times L$ square array with $L=4$ and $L=6$. In order to do so, we use the ramps shown in Fig.~\ref{fig:fig1}(b) with $\hbar \Omega_\text{max}/U = 2.3 $. This value is larger than the critical point $\left(\hbar\Omega/U\right)_c \approx 1.5$, such that the quantum phase transition line is crossed while ramping down $\Omega$, see Fig.~\ref{fig:fig1}(c). From the analysis of the final fluorescence images, we reconstruct the Rydberg density $n=\sum_i\ave{n_i}/N$, and the {\em connected} spin-spin correlation function
\begin{equation}
g^{(2)}(k,l)=\frac{1}{N_{k,l}}\sum_{(i,j)}[\ave{n_i n_j}-\ave{n_i}\ave{n_j}]\ ,
\label{eq:g2}
\end{equation}
where the sum runs over atom pairs $(i,j)$ whose separation is ${\bs r}_i-{\bs r}_j=(ka,la)$, and $N_{k,l}$ is the number of such atom pairs in the array~\footnote{In order to improve the statistics we symmetrize the experimental data for a given $(k,l)$ over the four quadrants ($\pm k,\pm l$) for the square lattice (and in an adapted way for the triangular lattice). Since these operations are symmetries of the setup, the symmetrization does not alter genuine features in the experimental data.}. For a perfect antiferromagnetic N\'eel ordering, $g^{(2)}$ takes the values $\pm1/4$ for $|k|+|l|$ even and odd, respectively.

Figure~\ref{fig:fig2}(a) shows  the spin-spin correlation function at the end of the ramp as a function of $x= \hbar\delta_{\rm final}/U$. We observe strong AF correlations, i.e. the sign of $g^{(2)}(k,l)$ changes according to the parity of the Manhattan distance $|k|+|l|$, in the region $0<x<4$ where AF order is expected, while the correlations vanish outside of this region. The amplitude of the AF correlations decreases with distance, in a way which is well captured by an exponential decay with a correlation length $\xi$, defined by $g^{(2)}(k,l)\propto (-1)^{|k|+|l|} \exp[-(|k|+|l|)/\xi]$, of about $1.5$ sites [see Figs.~\ref{fig:fig1}(d) and \ref{fig:fig3}(b)]. Repeating the same experiment  with a 1d chain yields the same correlation length (the particularity of the triangular lattice is discussed in Appendix~\ref{app:corrstriangular}).   Importantly, even though the correlation length is smaller than two sites for both the  1d chain and the square lattice, we are able to detect {\em finite}  correlations with the expected sign structure for up to five Manhattan shells, {\it i.e.}, almost over the whole array.

Another way to highlight AF correlations is to partition the array into the two N\'eel sublattices $A$ and $B$ and plot a two-dimensional histogram $P(n_A,n_B)$ with the $\ket{\uparrow}$ populations $n_A$ and $n_B$ of each sublattice as axes. For a perfect AF ordering, one would observe populations only in the two corners $(n_A=0,n_B=N/2)$ and $(n_A=N/2,n_B=0)$. The experimental results in Fig.~\ref{fig:fig2}(b) show that for $x=2.5$ (central plot), the sublattice population histogram is substantially elongated along the anti-diagonal $n_B=N/2-n_A$, which is not observed for $x<0$ or $x>4$. For comparison, Fig.~\ref{fig:fig2}(c) shows the corresponding histograms that would be obtained for an uncorrelated random state with the same average Rydberg density: there the elongation along the anti-diagonal is absent. 

In Fig.~\ref{fig:fig2}(d,e) we locate the boundaries of the AF phase. Panel (d) shows the mean Rydberg density $n$. For the $\Omega=0$ ground state of \eqref{eq:hising} in the NN limit with periodic boundary conditions, it should rise in steps, from 0 for $x<0$, to $1/2$ for $0<x<4$, and to 1 for $x>4$~\footnote{Using the $1/r^6$ interaction would lead to finer structures around $x=0$ and $x=4$, but with very narrow steps in $x$}. For open boundary conditions, additional steps are present in the $0<x<4$ region, see Appendix~\ref{app:openclusters}. Experimentally, the curve $n(x)$ varies continuously due to the finite duration of the sweep, as also observed in Refs.~\cite{Richerme2014,Schauss2015}. 
This mean density is therefore not a good observable to differentiate the paramagnetic and AF regions. Instead, we introduce the N\'eel structure factor $S_\text{N\'eel}$ to detect antiferromagnetic correlations in the AF region of the phase diagram:
\begin{equation}
    S_{\text{N\'eel}} = 4 \times \sum_{k,l} (-1)^{|k|+|l|}g^{(2)}(k,l)\ .
\label{eqn:structurefactor}
\end{equation}
This quantity is an estimator for the correlation volume, i.e., the number of spins correlated antiferromagnetically with a given spin. In a situation with true long range order, $S_{\text{N\'eel}}$ diverges linearly with the total ``volume" $N$ of the system, while it stays almost constant for short-range ordered correlations when the system sizes are larger than the correlation length. As shown in Fig.~\ref{fig:fig2}(e), this quantity indicates the presence of substantial short-range AF correlations for $0<x<4$, as expected from the phase diagram. 

The results presented so far demonstrate that by analyzing the correlation functions we can locate phase boundaries in the phase diagrams. In the next section, we will study how the correlations build up in time. 

\begin{figure*}[t]
\includegraphics[width=17.8cm]{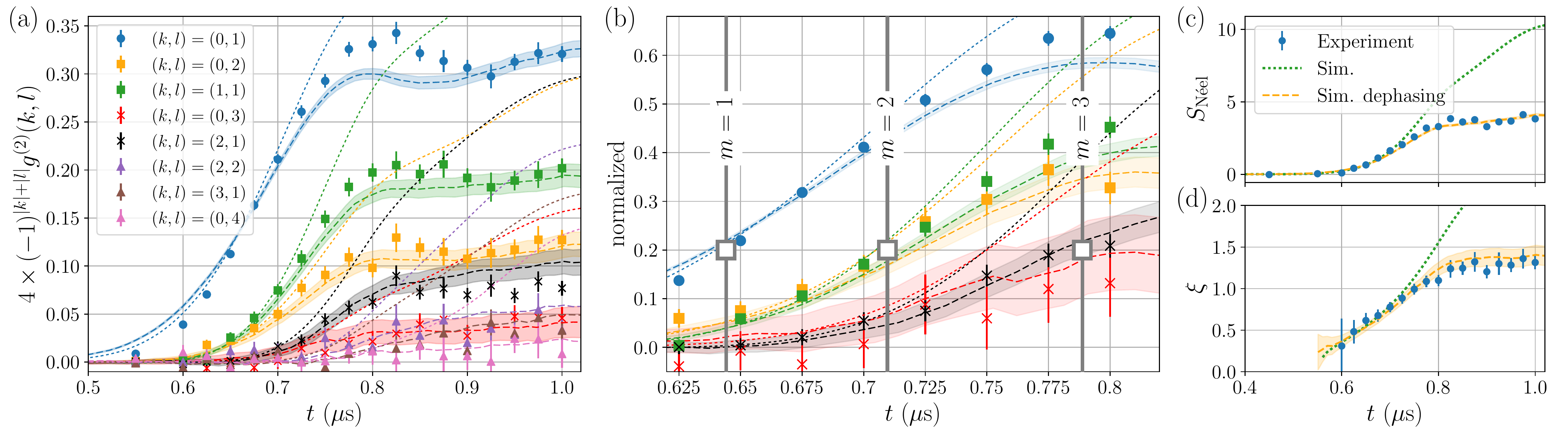}
\caption{Time evolution of a $6\times 6$ square array along the optimized ramp (see text). (a) Build-up of the correlation along the ramp. (b) Observation of a time delay between the build-up of significant correlations in increasing Manhattan shells $m$. This is a manifestation of a Lieb-Robinson type limitation in the build-up of correlations between distant sites (see text). (c) N\'eel structure factor and, (d) correlation length $\xi$. All figures: the dotted (dashed) lines correspond to the result of a numerical exact diagonalization simulation for a $4\times4$ lattice without (with) dephasing.}
\label{fig:fig4}
\end{figure*}

\section{Exploring the time- and space-dependence of correlations}\label{Sec:time_space_dependence}

In the following we first investigate in Sec.~\ref{Ssec:varying_speed} how $S_\text{N\'eel}$ depends on the ramp speed and find a duration $t_{\rm tot}$ that maximizes its value. Then, with these settings, we study in Sec.~\ref{Ssec:time_dependence} the temporal build-up of correlations during the optimized sweep and observe delays between the growth of correlations at increasing distances. Eventually, in Sec.~\ref{Ssec:space_dependence}, we analyze the spatial structure of these correlations and give a qualitative explanation of the observed patterns in terms of a short-time expansion.

\subsection{Correlations after ramps of varying durations}\label{Ssec:varying_speed}

We perform experiments on a 36-site square lattice using the sweep shown in \figref{fig:fig1}(b) with fixed parameters $\hbar\delta_{\text{final}}/U = 1.7$, $\hbar\Omega_{\text{max}}/U = 0.7 < \left(\hbar\Omega/U\right)_c$. In contrast to the previous experiment, the quantum phase transition line is crossed while ramping the detuning $\delta$ and we vary the duration of the sweep $t_\text{sweep}$ (and therefore $t_{\rm tot}$). The initial parameters are chosen such that the ground state at $t=0$ corresponds to $\ket{\downarrow \downarrow \cdots}$ (no Rydberg excitation) and the ground state at the final parameters is an AF~\footnote{In the bulk or for $L$ even, the AF state is the equal superposition of the two classical N\'eel states. It has a uniform Rydberg density $n_i=1/2$, while $g^{(2)}(k,l)=(-1)^{|k|+|l|} \times 1/4$. In stark contrast, on open boundary $L\times L$ arrays with $L$ odd, a {\em single} N\'eel ground state is selected. This state exhibits a staggered pattern of Rydberg densities, while the $g^{(2)}$ correlations vanish identically.}.

Fig.~\ref{fig:fig3}(a) shows two-dimensional plots of the experimental $g^{(2)}$ correlation functions for four different values of $t_{\text{tot}}$. For the shortest sweeps, the correlations remain weak. For intermediate durations ($t_{ \text{tot}}\sim 1 \, \mu$s) strong correlations emerge, with a staggered structure extending over many Manhattan shells, as shown in Fig.~\ref{fig:fig3}(b). For even larger sweep durations, the correlation signal decreases with increasing sweep duration. In Fig.~\ref{fig:fig3}(c) we show the value of $S_\text{N\'eel}$ as a function of $t_{ \text{tot}}$: starting from small values for short ramps, it exhibits a broad maximum for $t_{ \text{tot}}\sim 1 \, \mu$s and slowly decreases for longer ramp durations.

In order to gain a better understanding of the behavior described above, we compare the data with numerical simulations of the dynamics of the system governed by the Hamiltonian of \eqref{eq:hising} with the full van der Waals interactions for the sweeps used in the experiment~\footnote{We have checked that using NN interactions instead of the exact van der Waals coupling does not lead to appreciable changes in the dynamics.}. We show in \figref{fig:fig3}(c) the results of the simulation on a $4\times4$ square lattice. The green line corresponds to a unitary evolution which models well the experiment only for short sweep durations. The dashed yellow line includes the decoherence observed in the experiment on the excitation of one atom through an empirical dephasing model (see Appendix~\ref{app:dissipation}). We observe a remarkable agreement between this local ({\it i.e.} single atom) dephasing model with no adjustable parameters and the experiment over a large range of $t_{ \text{tot}}$. This indicates that the saturation and the decay of the correlations for longer sweeps is due to (single-particle) decoherence. 

We now analyze the spin-spin correlations for different Manhattan distances $|k|+|l|$. In \figref{fig:fig3}(d) we observe the build up of correlations up to the fourth shell $|k|+|l|=4$, all of them being antiferromagnetically staggered, $g^{(2)}(k,l) \propto (-1)^{|k|+|l|}$. For short sweeps the correlations sharply increase with increasing duration, saturate for longer sweeps, and decay, again due to decoherence. The simulation including dephasing (dashed lines) reproduces well this trend. 

\subsection{Build-up of correlations along a ramp of fixed duration}\label{Ssec:time_dependence}

\begin{figure*}[t]
\centering
\includegraphics[width=\linewidth]{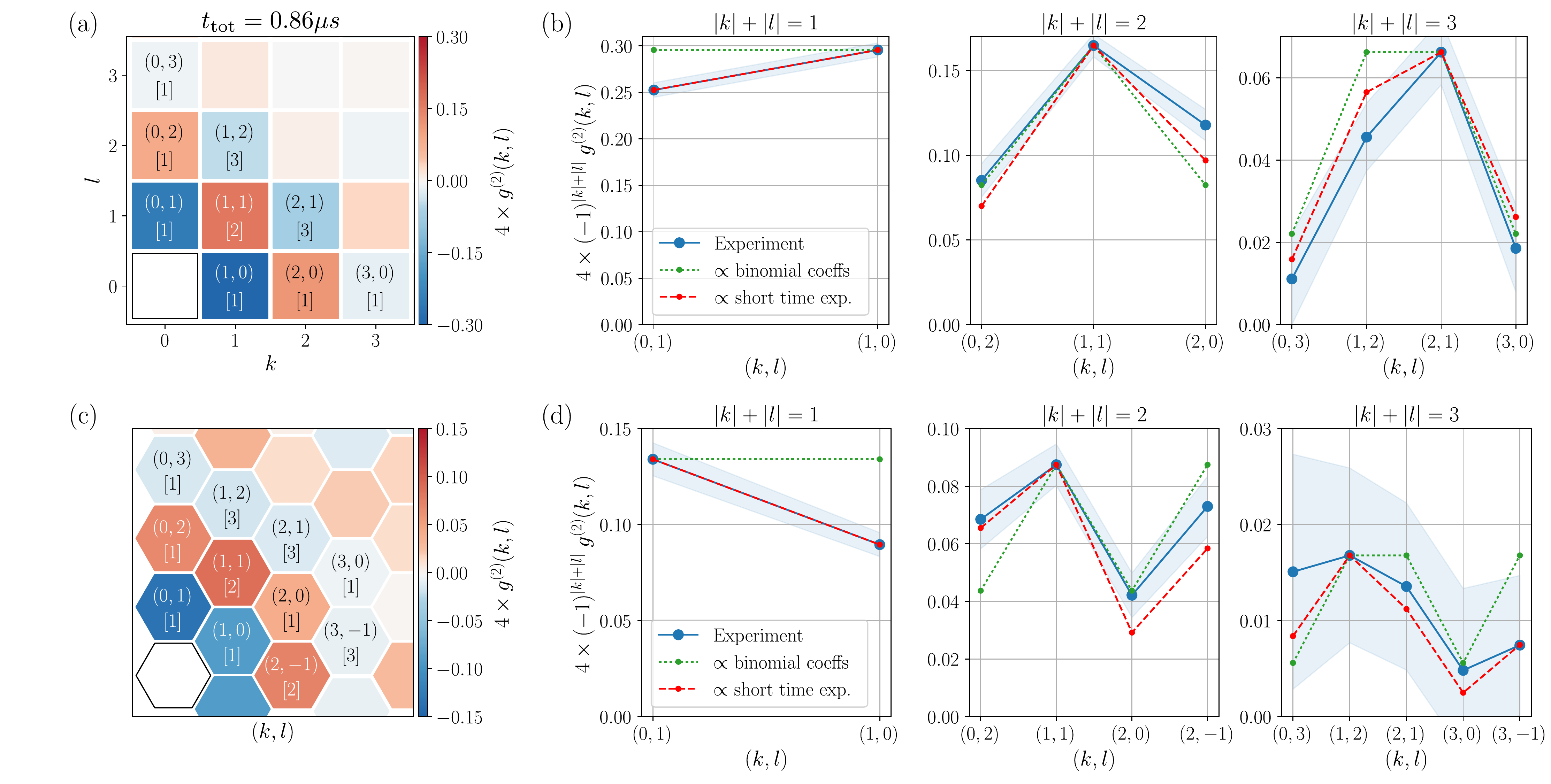}
\caption{Spatial structure of the spin-spin correlations on the 36-site square (a-b) and on the 36-site triangular lattice (c-d). Experimental data after a sweep is shown in (a) and (c) together with the values of $(k,l)$ for the first three shells $|k|+|l|$ (see~\figref{fig:fig1}(d) for the full range). The numbers in brackets $[c]$ give the number of linking paths contributing to the short-time expansion. (b) and (d) dependence of the correlation on $(k,l)$ for several Manhattan distances with a qualitative comparison to binomial coefficients (green dotted lines) and leading-order short-time expansion (red dashed lines), see text for details. The shaded region corresponds to the s.e.m. of the experimental data.}
\label{fig:fig5}
\end{figure*}

In a subsequent experiment, we analyze how the AF correlations build up in time, during the sweep that maximizes $S_\text{N\'eel}$ ($t_\text{tot} = 1.0 \, \mu$s) identified in the previous subsection. Stopping the dynamics after a variable time $0<t<t_{\rm tot}$, Fig.~\ref{fig:fig4} shows the time evolution of the correlations after entering the AF region identified in Fig.~\ref{fig:fig2} at $\delta(t) > 0$ ($t > 0.5 \, \mu$s). We observe that most of the correlations build up for $0.6 <t< 0.8 \, \mu$s and then freeze at larger times. When comparing again to the numerical simulation of the dynamics including the single-particle dephasing, we obtain a remarkable agreement with the data. 

Fig.~\ref{fig:fig4} also features a striking effect: we observe a delay in the build-up of correlations between the different Manhattan shells $m = |k|+|l|$, highlighting the finite speed for the spread of correlations. To quantify this effect we normalize the correlations for each distance such that the corresponding dephasing-free simulation reaches a maximum of 1 at large time [\figref{fig:fig4}(b)]. Fixing an arbitrary threshold of $0.2$, we observe that the nearest neighbor shell $(m=1)$ reaches this threshold at $t \approx 0.64 \, \mu$s, the second shell $(m=2)$ at $t\approx 0.71 \, \mu$s and finally the third shell $(m=3)$ at $t\approx 0.79 \, \mu$s (see gray vertical lines). This delay is a manifestation of the finite propagation speed of correlations as theoretically predicted by Lieb-Robinson bounds~\cite{Lieb1972,Bravyi2006}. Such bounds are well explored in the context of quench dynamics~\cite{Laeuchli2008,Cheneau2012,Langen2013,Richerme2014,Jurcevic2014}, but their importance for state preparation protocols is not equally well known (see Appendix~\ref{app:liebrobinson} for a brief review on Lieb-Robinson bounds adapted for the present context).

Fig.~\ref{fig:fig4}(b) also reveals that when the correlations on the first shell reach the threshold, correlations for higher Manhattan distances are suppressed, but still detectable for $m = 2$. This illustrates the known fact that the correlations outside the Lieb-Robinson cone are finite, albeit exponentially suppressed (see also Appendix~\ref{app:liebrobinson}). To recover this fact, we introduce a simple analytical approach based on a short-time expansion method (see Appendix~\ref{app:short:time} for details and specific results). We find analytical expressions in powers of the duration $T$ of the ramp for the connected $g^{(2)}$ correlation functions for different Manhattan distance $m$ valid in the limit ($UT/\hbar,\Omega T,\delta T \ll 1$). We obtain nearest-neighbor correlations ($m=1$) of order $T^6$ and next-nearest neighbor correlations ($m=2$) of order $T^{10}$. More generally, the leading order of the expansion for two sites separated by a  Manhattan distance $m$ appears at order $T^{2+4m}$ thus suggesting that at a given (short) time the correlation decreases exponentially with $m$ (see more details in Appendix~\ref{app:short:time}). This scaling explains qualitatively the observed time dependence in Figs.~\ref{fig:fig4}(a) and (b), even though strictly speaking the range of applicability of the short-time expansion is limited to times much shorter than those shown there (for which the correlations would still be extremely small, and thus very challenging to measure experimentally). However, our numerical exact diagonalization results without and with dissipation as well as the experiment show that the qualitative features of the short-time expansion prevail for the considered non-perturbative times and the actual ramp shapes.

\subsection{Build-up of spatial structures on the square and triangular lattices}\label{Ssec:space_dependence}

We finally analyze the spatial structure of the correlations in more detail. Both Fig.~\ref{fig:fig3}(d) and Fig.~\ref{fig:fig4}(a,b) show that the correlations do not depend only on the Manhattan distance $m = |k|+|l|$, but also, for fixed $m$, on $k-l$. For instance, for second neighbors ($m=2$), the correlations for $(k,l)=(0,2)$ or $(2,0)$ is about half of those along the diagonal $(k,l)=(1,1)$. This spatial structure, absent in a 1d setting, has not, to our knowledge, been experimentally observed in 2d.

The observed spatial structures in the correlations within a given Manhattan shell $m$ can also be captured by the short-time expansion. The leading order coefficient of a given correlator $g^{(2)}(k,l)$ depends on the number of paths on the lattice linking the two sites as detailed in Appendix~\ref{app:short:time:embedding}. Fig.~\ref{fig:fig5} shows a comparison of the spatial structure in experimental data for both a square and a triangular lattice after a ramp of finite duration. In panels (a) and (c), the number of linking paths is given for $k\ge l\ge 0$ by the binomial coefficient $C^{m}_{k-l}$ for both lattices and is shown in brackets. In panels (b) and (d) we analyze the correlations on the square and triangular lattice. The green dotted lines show the $k-l$ dependence of $g^{(2)}(k,l)$ assumed to be given only by the binomial coefficients, normalized to the maximal correlation value for each subplot. 

The precision of the experiment even allows us to  observe a small asymmetry in each shell $m={\rm const.}$, due to a slight residual anisotropy of the interaction. An extension of the short-time expansion to the anisotropic case yields an analytical expression from which we extract the ratio $U_z/U_w$ of the nearest-neighbor interactions. We use this interaction anisotropy in the analytical expression to obtain the correlation for larger $|k|+|l|$. The results are shown as red dashed lines. 

Interestingly, for triangular lattices, in contrast with the case of square arrays, the spatial structures of the correlations that one observes experimentally, and which are well reproduced by the short-time expansion, do not reflect directly those one would obtain in the ground state (see Appendix~\ref{app:corrstriangular}). A detailed study of this specific behavior, which may be a dynamical signature of the frustrated character of the triangular geometry, is however beyond the scope of the present work.

\section{Conclusion and Outlook}

In conclusion, we have used a Rydberg-based platform to study antiferromagnetic correlations in a synthetic Ising magnet with different geometries. Using dynamical variations of the parameters, we explored the phase diagram and prepared arrays exhibiting antiferromagnetic order with pronounced Manhattan structures. We also studied the growth of the correlations during the sweeps of the parameters. We observed delays in the build-up of correlations between sites at different distances, a feature linked to the Lieb-Robinson bounds for the propagation of correlations in a system with nearest-neighbor interactions. We were able to understand the spatial structure of the correlations after short to intermediate evolution times using an analytical short-time expansion of the evolution operator. Finally, we obtained remarkable agreement between the data on the dynamics of the correlations
and numerical simulations using a local ({\it i.e.}~single particle) dephasing model. 

In the future, the time over which coherent simulation can be performed in such artificial quantum magnets could be extended by a better understanding of the dephasing mechanisms at play in the coherent excitation of Rydberg states \cite{RabiPaper}. We will then be able to explore geometric frustration on triangular or Kagome lattices \cite{Humeniuk2016,Laeuchli2015}. Another promising avenue is the study of magnets described by the XY model implemented using resonant dipole-dipole interactions~\cite{Barredo2015,deLeseleuc2017}: there the coherent drive is obtained by using microwaves, and much lower dephasing rates are expected; moreover the interaction decays slowly, as $1/r^3$, and long-range effects should be more prominent \cite{Peter2012}. 

{\em Note:} Similar antiferromagnetic correlations in 2d have been observed recently at Princeton University, using a system of Li Rydberg atoms in a quantum gas microscope.

\begin{acknowledgments}
This work benefited from financial support by the EU [H2020 FET-PROACT Project RySQ], by the ``Investissements d'Avenir'' LabEx PALM (ANR-10-LABX-0039-PALM, projects QUANTICA and XYLOS) and by the R\'egion \^Ile-de-France in the framework of DIM Nano-K. MS, LPH and AML acknowledge support by the Austrian Science Fund for project SFB FoQus (F-4018). The computational results presented have been achieved in part using the Vienna Scientific Cluster (VSC). This work was supported by the Austrian Ministry of Science BMWF as part of the UniInfrastrukturprogramm of the Focal Point Scientific Computing at the University of Innsbruck. 
\end{acknowledgments}

$^{\ast}$ V.~L., S. de L. and M. S. contributed equally to this work.

\appendix

\section{Experimental details}
\label{app:exp}

Here we give more details about our experimental setup and on the mapping of the Rydberg system onto an Ising antiferromagnet.

We use a two-photon excitation scheme to excite the atoms to the Rydberg state. The two lasers at 795~nm and 475~nm, with corresponding Rabi frequencies $\Omega_{\rm r}$ and $\Omega_{\rm b}$, and a detuning $\Delta\simeq 2\pi\times 740$~MHz from the intermediate $\ket{5P_{1/2},F=2}$ state result in an effective Rabi frequency $\Omega=\Omega_{\rm r}\Omega_{\rm b}/(2\Delta)$ for the coupling between $\ket{\downarrow}$ and $\ket{\uparrow}$, but also introduce lightshifts $(\Omega_{\rm r}^2-\Omega_{\rm b}^2)/(4\Delta)$ that add up to the two-photon detuning $\delta$. In order to generate the ramps shown in Fig.~\ref{fig:fig1}(b), we thus compensate for these additional lightshifts. An AOM is used for changing dynamically the amplitude and frequency of the red beam. Due to the finite size of our excitation beams, the atoms do not all experience exactly the same $\delta$ and $\Omega$. For our arrays with a size $\sim 40\;\mu$m, the inhomogeneities of $\Omega$ are below 15$\%$, while the detuning does not differ from its value on the central atom by more than 150~kHz.

\begin{figure}[t]
\includegraphics[width=7cm]{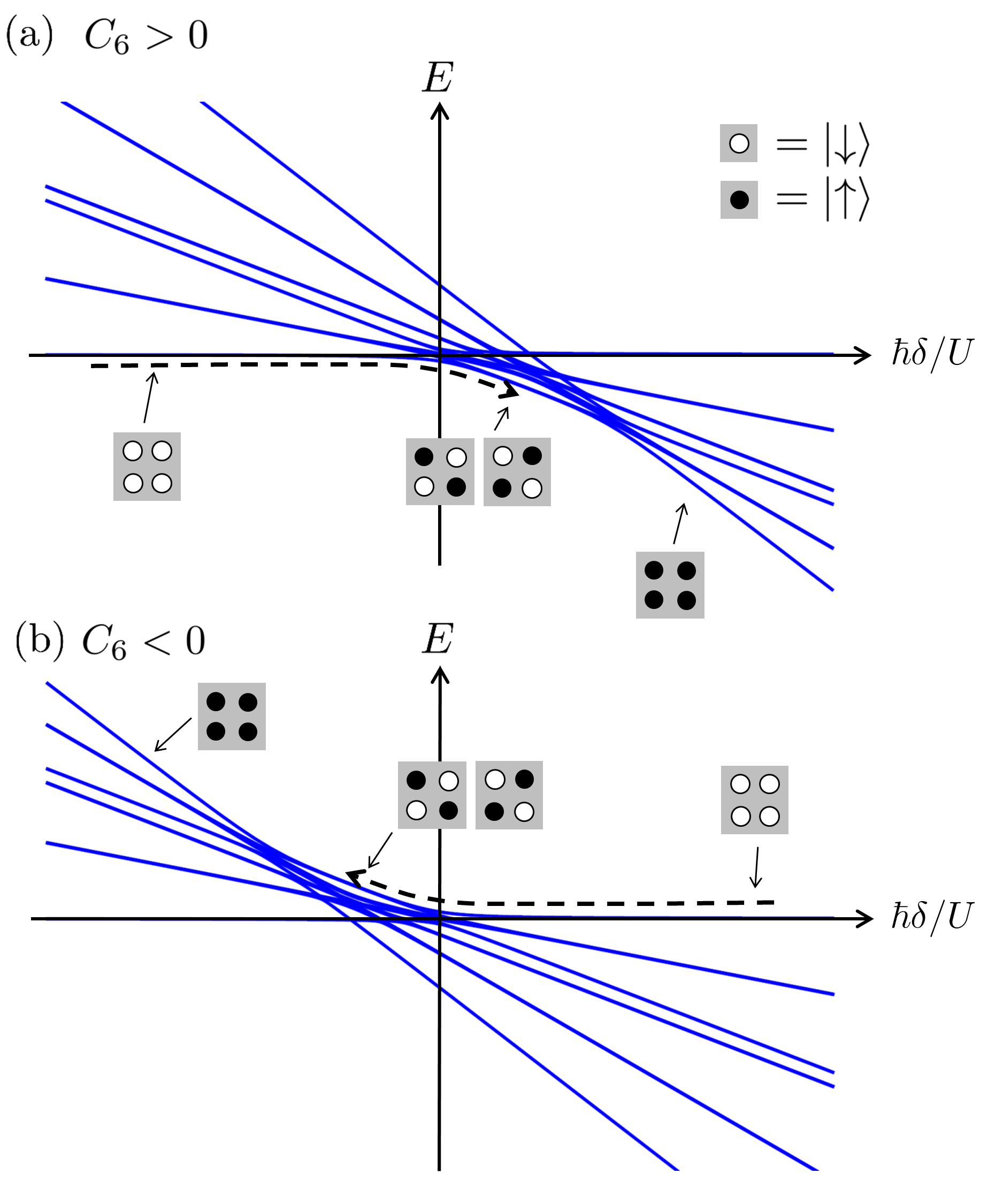}
\caption{Influence of the sign of $C_6$ on the many-body spectrum. Eigenenergies of a small $2\times2$ square system as a function of $\delta$ for $\hbar\Omega/U\simeq0.5$. 
(a) For a repulsive interaction $C_6>0$, the ground state of $H$ is antiferromagnetic for $0<\hbar\delta<2U$. It can be reached from $\ket{\downarrow}^{\otimes 4}$ 
by adiabatically following the ground state starting from negative values of $\delta$ 
(trajectory in dashed arrows). (b) For an attractive interaction $C_6<0$, the antiferromagnetic state is the most excited state, and the sign of the detuning 
needed to reach it from $\ket{\downarrow}^{\otimes 4}$ is reversed.}
\label{fig:signC6}
\end{figure}

The detection of the state of each atom relies on the loss of atoms in the Rydberg state, which are not recaptured in the optical tweezers. This detection method is thus subject to small detection errors \cite{RabiPaper}. In particular, an atom actually in $\ket{\downarrow}$ has a small probability $\varepsilon$ to be lost and thus incorrectly inferred to be in $\ket{\uparrow}$. As in \cite{Labuhn2016}, we measure $\varepsilon$ in a calibration experiment, and then include the effect of detection errors on all the theoretical curves. 

An \emph{antiferromagnetic} phase is expected when interaction between parallel spins are repulsive, corresponding to the case where $U>0$ in the Hamiltonian Eq.~(\ref{eq:hising}). Fig.~\ref{fig:signC6}~(a) shows the energy levels for a 2$\times$2 square matrix of atoms as a function of the detuning $\delta$, with $U>0$ and $\hbar \Omega \simeq 0.5\,U$. The lowest energy configuration with two $\ket{\uparrow}$ spins is then an antiferromagnetic state for $0<\hbar\delta<2U$. Then, starting from a detuning $\delta < 0$ with all atoms in $\ket{\downarrow}$, and ramping it up to $\delta \approx U/\hbar$ (dashed arrow), the lowest energy state is evolving adiabatically to the antiferromagnetic state. In our experiment, the interaction between $\ket{\uparrow}=\ket{64D_{3/2},m_j=3/2}$ spin states are in fact attractive ($U<0$). Consequently, the antiferromagnetic state is never the lowest energy configuration whatever the value of $\delta$, see Fig.~\ref{fig:signC6}~(b). Nevertheless, starting from a detuning $\delta > 0$ and ramping it down to $\delta \approx U/\hbar$, the system evolves from the initial state to an antiferromagnetic state while following the highest-energy level. In this case, preparing the antiferromagnetic state actually means obtaining the ground state of $-H$, hence the change of sign of the detunings needed to reach the correlated phase. For our parameters, the van der Waals interaction is not only attractive, but it is also slightly anisotropic \cite{Barredo2014,Browaeys2016,deLeseleuc2017b}, being about three times as small in the horizontal direction as in the vertical one (along $\hat{z}$). We compensated for this difference by slightly distorting the arrays along $\hat{z}$ by a factor $\sim 3^{1/6}$.

\begin{table}[t!]
\begin{center}
\begin{tabular}{lcccccccc}
\hline
{\bf Figure} & $U/h$ &  $\Omega_\text{max}/(2\pi)$ & $\delta_\text{final}/(2\pi)$ & & $t_\text{rise}$ & $t_ \text{sweep}$ & $t_\text{fall}$\\
             & (MHz) & (MHz) & (MHz) && ($\mu$s) & ($\mu$s) & ($\mu$s) \\
\hline \hline
& & & & & & & \\ 
$2$ & $1.0$ & $2.3$ & $[-2,6]$ & & $0.25$ & $\displaystyle{\frac{\delta_{\text{final}}-\delta_0}{2\pi\cdot 10 \, ({\rm MHz})}}$ & $0.5$ \\
& & & & & & & \\ 
$3$ & $2.7$ & $1.8$ & $4.5$ & & $0.25$ & $[0.1,1.3]$ & $0.25$  \\
& & & & & & & \\ 
$4$ & $2.7$ & $1.8$ & $4.5$ & & $0.25$ & $0.44$ & $0.25$  \\
& & & & & & & \\ 
$5$(a) & $2.7$ & $1.8$ & $4.5$ & & $0.25$ & $0.44$ & $0.25$  \\
$5$(b) & $0.8$ & $0.6$ & $1.6$ & & $0.25$ & $5.5$ & $0.25$  \\
& & & & & & & \\ 
\hline
\end{tabular} 
\caption{Experimental parameters used for the data presented in the main text.}
\label{tab:tab1}
\end{center}
\end{table}

\section{Description of the ``order-by-disorder'' process}
\label{app:orderbydisorder}

As mentioned in Sec.~\ref{Sec:phase_diag}, the triangular lattice presents an interesting feature for $\hbar\delta/U=z/2$: there the model is fully frustrated in the classical limit; the interactions on the bonds of each single triangle of the lattice cannot be simultaneously satisfied under the given density constraint $n=1/2$ leading to an extensive ground-state degeneracy~\cite{Wannier1950}. Upon switching on the transverse field, this degeneracy is lifted by a process called ``order-by-disorder'' (OBD), i.e.~the addition of ``disorder'' (here the quantum fluctuations provided by the transverse field $\Omega$) selects a subset of configurations displaying an ordered pattern. In the case under consideration here, OBD leads to the selection of a 3-sublattice ordered phase with Rydberg occupations of the three sublattices $(n_A,n_B,n_C) = (\lambda,1-\lambda,1/2), \lambda \neq 1/2$. This phase later undergoes a quantum phase transition in the 3d-XY universality class to the paramagnetic phase for a larger $\Omega$~\cite{Moessner2000,Moessner2001}.

\section{Finite-size effects}
\label{app:openclusters}

\begin{figure}[t]
\includegraphics[width=8.6cm]{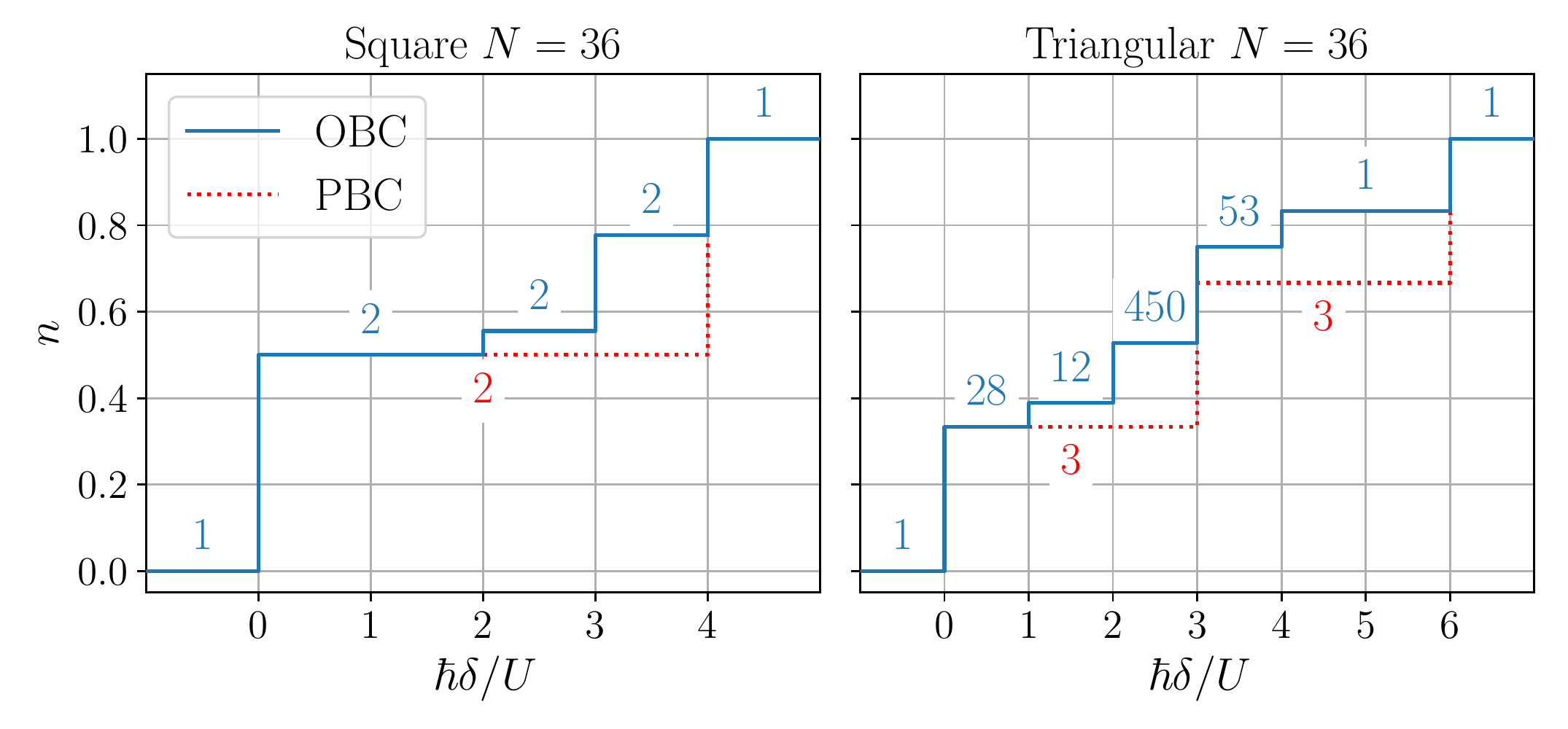}
\caption{Classical ($\Omega=0$) phase diagrams for arrays with open boundary conditions. Rydberg density $n$ as a function of $x=\hbar\delta/U$ for the $N=36$ square 
and triangular arrays used in the experiment. Blue lines correspond to open boundary conditions while red dashed lines correspond to periodic boundary conditions (bulk phase diagram). The numbers indicate the degeneracy of the classical configurations for the distinct plateaus.}
\label{fig:OBC}
\end{figure}

The two-dimensional arrays used here present open boundary conditions (OBC). The lattice sites along the boundary have less neighbors than the sites in the bulk and therefore experience less interactions. This modifies the ``phase diagram'' with respect to the bulk phase diagrams presented in \figref{fig:fig1}. In \figref{fig:OBC} we show the classical (i.e.~$\Omega=0$) ground-state configurations for the $N=36$ square and triangular arrays used in the experiment and compare them to the bulk phase diagrams. The sites at the boundary are more easily excited to the Rydberg state when $\delta$ increases, leading to a larger number of density plateaus. We have checked that for the parameters
used in this paper the OBC and bulk phases feature the same typical short-range correlations.

The N\'eel structure factor $S_\text{N\'eel}$ shows almost no finite-size effects for short-range ordered systems as long as the system sizes are larger than the correlation length. 
For finite arrays the number of pairs of sites with larger distances $|k|+|l|$ is strongly reduced leading to larger errors for the connected correlations $g^{(2)}(k,l)$ due to reduced statistics.  In order to keep reasonable values and errors for the N\'eel  structure factor we have restricted the summation to $|k|+|l|\leq4$ in the experimental evaluation of $S_\text{N\'eel}$.

\section{Numerical Methods}
\label{app:numerics}

We perform numerical exact diagonalization simulations of the time evolution on medium-size lattices for both Hamiltonian systems without any dissipation and the same systems 
with additional dephasing terms in the framework of a master equation in Lindblad form. In both cases we construct the complete many-body Hilbert space of the system on the finite lattice and do not perform any truncation to calculate the time evolution. Therefore, during the time evolution the system can explore the entire Hilbert space and all phases in the phase diagram can potentially be obtained. 

\subsection{Unitary time evolution}
To simulate the unitary time evolution of a time-dependent Hamiltonian $H(t)$, we approximate $H(t)$ by a piecewise constant Hamiltonian $H_{\text{approx}}(t)$ 
on $n_{\text{steps}}$ intervals of length $\Delta t$ such that
\begin{equation}
H_{\text{approx}} (t) = \frac{H(n \Delta t) + H((n+1) \Delta t)}{2}, 
\end{equation}
for $n \Delta t \leq t < (n+1) \Delta t$. We then compute the evolution operator within each interval with a  Krylov-type matrix exponential approach~\cite{Park1986} 
and use the evolved state of the previous interval as starting state. We choose $n_{\text{steps}}$ such that the results do not differ from a simulation with $n_{\text{steps}}/2$ intervals within a demanded accuracy. In most of the presented simulations $n_{\text{steps}}=200$. 

We are able to calculate the unitary time evolution of system sizes of up to
about 30 lattice sites with a reasonable amount of resources. In the present
study most of the numerical results have been obtained on $4\times4$ and
$5\times5$ square lattices. Due to the relatively short correlation lengths
observed in the experiments, the numerical results on the smaller systems can
nevertheless be used to model the observed correlation functions on the larger,
experimentally realizable lattices.

\subsection{Time evolution in the presence of dephasing}
\label{app:dissipation}

The experimental system features several sources of imperfections~\cite{RabiPaper} and the description of its evolution as a pure Hamiltonian is not exact. In particular, phase noise of the excitation lasers and Doppler shifts lead to dephasing, already for a single particle. We thus perform simulations with a phenomenological local dephasing model with a rate $\gamma$ obtained by fitting experimental single atom Rabi oscillations. We obtain dephasing rates $\hbar\gamma/U \approx 1.1 -1.4$ for the data shown in this paper.
The time evolution of the density matrix $\rho(t)$ of the many-body system is then described by the following master equation in Lindblad form
\begin{equation}
    \frac{\dd}{\dd t} \rho = -\frac{i}{\hbar} \left[H,\rho\right] + \mathcal{L}[\rho]\, ,
    \label{eq:mastereq}
\end{equation}
with a Liouvillian
\begin{equation}
    \mathcal{L}[\rho] = \sum_i \frac{\gamma}{2} \left( 2 n_i \rho n_i - n_i \rho - \rho n_i \right)\,.
\end{equation}

Direct simulations of the master equation are only possible for small lattices as the memory demand for saving density matrices grows as $\mathcal{O}(4^N)$ for the considered Hilbert space on $N$ sites. Thus we use a Monte Carlo wave-function (MCWF) method~\cite{Dalibard1992,Dum1992,Molmer1993} where only a single wave-function has to be stored such that one can simulate system sizes of up to around twenty sites. The MCWF method evolves a starting state with an effective non-hermitian Hamiltonian; at each time step a quantum jump corresponding to the given dissipation operator collapses the evolved state with a given probability. Averaging over many such quantum trajectories ---we use 1280 for the presented results--- allows reconstructing the density matrix and thus computing any observable during the time evolution. In practice, to perform simulations with dissipation, we use the Python toolbox ``QuTiP''~\cite{Johansson2013}.

\section{Lieb-Robinson bounds}
\label{app:liebrobinson}

\begin{figure}[t]
\includegraphics[width=8.6cm]{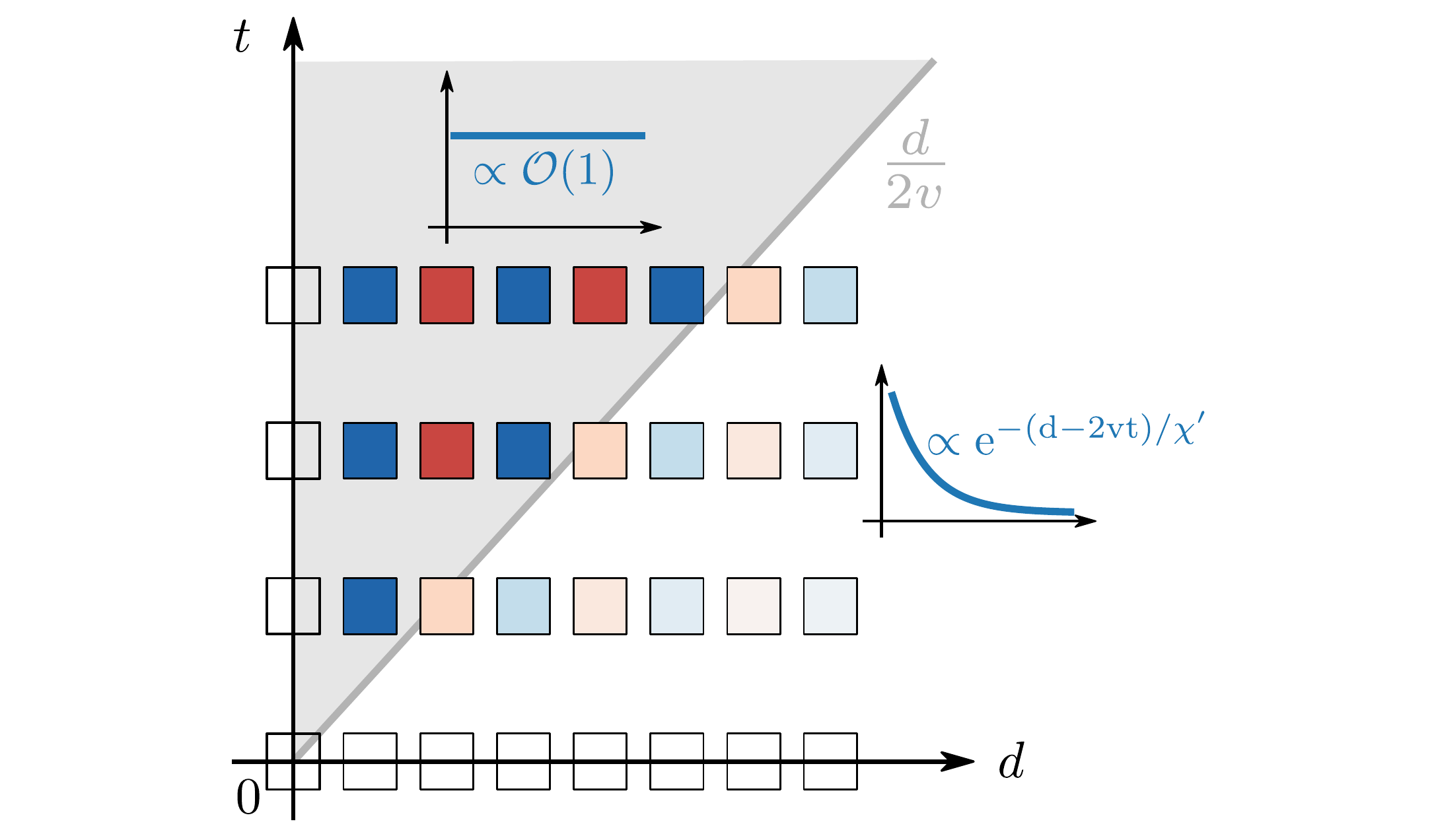}
    \caption{Illustration of the Lieb-Robinson bounds. ``Significant'' correlations spread along a light-cone $d=2\,v\,t$. Along and inside the light-cone (grey shaded area) correlations of order $\mathcal{O}(1)$ can develop. Outside the light-cone $d>2\,v\,t$ correlations are immediately built up for each $t>0$ but are exponentially suppressed in $d-2\,v\,t$.}
\label{fig:lieb_robinson}
\end{figure}

In non-relativistic quantum mechanics, one might believe that information propagation is instantaneous because  there is no {\em explicit} speed of light limiting the propagation. However, in 1972 Lieb and Robinson~\cite{Lieb1972} proved that in an extended quantum many-body system with a finite-dimensional local Hilbert space and sufficiently local interactions, there is nevertheless a characteristic velocity emerging, which defines an approximate light-cone for the propagation of information and implementing causality. 

In Ref.~\cite{Bravyi2006} the Lieb-Robinson bound has been generalized to equal-time connected correlation functions of two operators (normalized to unity) acting on spatial regions $A$ and $B$ (of size $|A|$ and $|B|$)  at a distance $d$ apart from each other after some time evolution of duration $t$ and starting from an initial state with exponentially decaying correlations with a correlation length $\chi$. Then the connected equal-time correlation $g^{(2)}(d,t) := \langle O_A(t) O_B(t) \rangle- \langle O_A(t) \rangle\langle O_B(t) \rangle$ is bounded as follows:
\begin{equation}
|g^{(2)}(d,t)| \leq \bar{c}(|A|+|B|)\exp[-(d-2vt)/\chi'],
\end{equation}
with $\chi'=\chi+2\zeta$. The coefficients $\bar{c},v$ and $\zeta$ depend on the Hamiltonian and the considered operators $O_A$ and $O_B$. The velocity $v$ is particularly important and is called the Lieb-Robinson velocity.

The importance of this result for the present study is that correlations at a distance $d$ are exponentially suppressed in  $(d-2vt)/\chi'$ as long as the time $t<d/(2v)$. After that time, the correlations are bounded by a number of $\mathcal{O}(1)$. This time dependence is visualized in Fig.~\ref{fig:lieb_robinson}. In closing we note that the theorem proves rigorous theoretical bounds, however the actual dynamics does not necessarily saturate these bounds.

\begin{figure}[t]
\includegraphics[width=8cm]{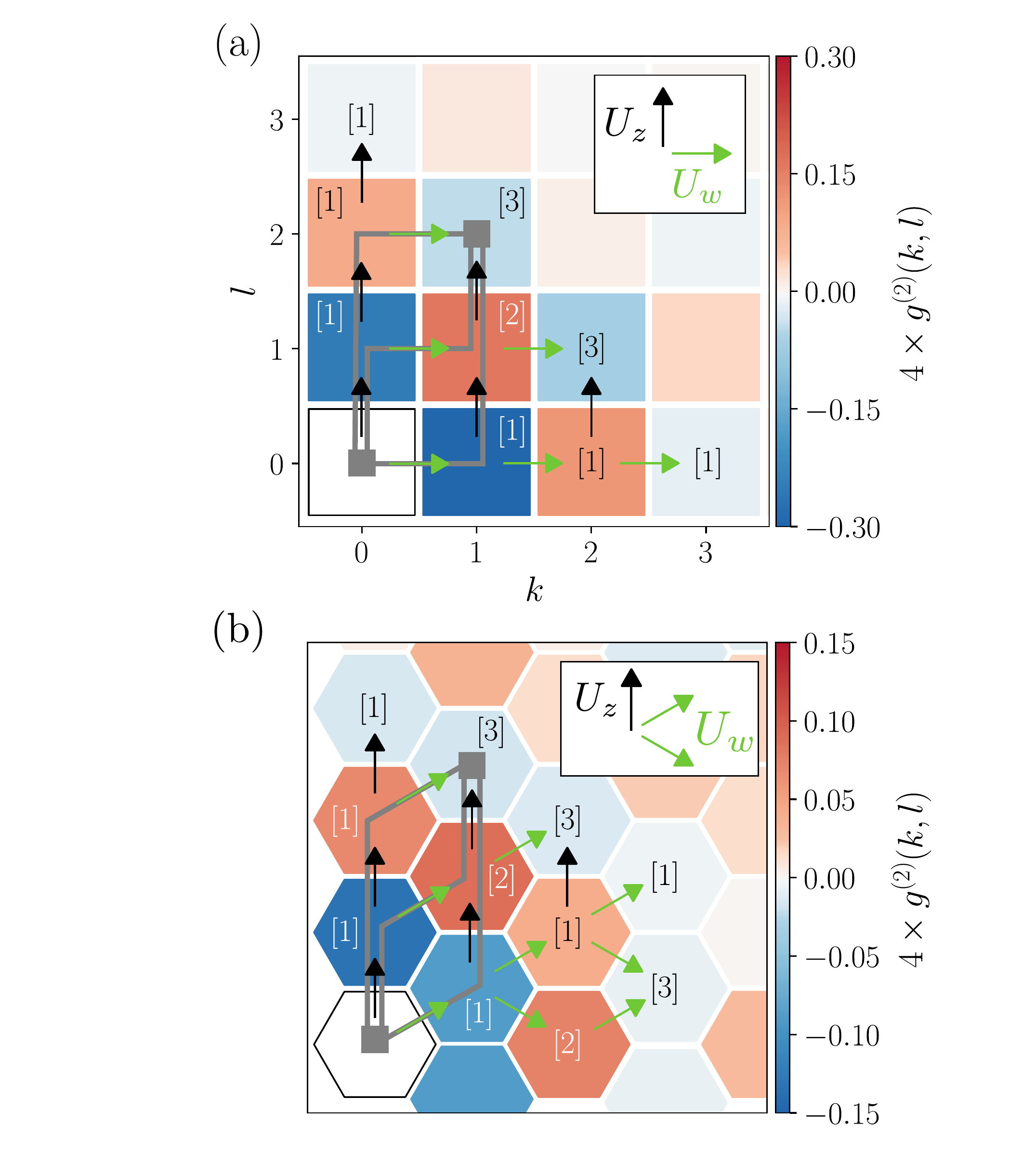}
\caption{Illustration of the different paths connecting two sites with a distance $(k,l)$ contribution to the leading-order short-time expansion on the (a) square lattice and (b) triangular lattice. The number in brackets shows the binomial coefficient $C^{|k|+|l|}_{|k-l|}$ (assuming $k,l\ge 0$) corresponding to the number of distinct paths. The arrows show the potentially different couplings along paths contributing to the short-time expansion. The grey lines exemplary show the three distinct paths contributing
to the correlation indicated by a grey square.}
\label{fig:binomialcoeffs}
\end{figure}

\section{Short-time expansion}
\label{app:short:time}
In this appendix we provide more details on the short-time 
expansion of the connected correlation function $g^{(2)}$.

\subsection{Theoretical description}
In order to simplify the notation we use the nearest-neighbor Hamiltonian ($\hbar=1$):
\begin{equation}
H(t)=\sum_{\langle i,j\rangle} U_{ij}\ n_i n_j - \delta(t) \sum_i n_i + \Omega/2 \sum_i \sigma^x_i\ .
\label{eqn:simpleham}
\end{equation}
Here, we keep the possibility that nearest-neighbor interactions 
could be different from atom to atom. 
We consider simple ramp shapes characterized by a constant $\Omega$, a duration $T$
and a linear $\delta(t)$-dependence 
between $\delta(0)=\delta_0$ and $\delta(T)=\delta_\text{final}$.

We are interested in the limit of very short duration $T$ and 
in particular how the correlations at various 
distances build up as we vary the duration $T$. 
Formally, we compute the full many-body propagator $\hat{U}(T)$ which solves the time-dependent Schr\"odinger equation
for the time-dependent Hamiltonian~\eqref{eqn:simpleham}, allowing to determine the many-body wave function at any time $T$ via 
$|\psi(T)\rangle=\hat{U}(T)|\psi(0)\rangle$. We then express the connected correlation function as 
\begin{eqnarray}
\label{eqn:appg2}
g^{(2)}[(k,l)](T)&=&\langle\psi(T)|n_{(0,0)} n_{(k,l)}|\psi(T)\rangle\\
&-&\langle\psi(T)|n_{(0,0)}|\psi(T)\rangle\langle\psi(T)|n_{(k,l)}|\psi(T)\rangle\nonumber
\end{eqnarray}

For the linear ramps  considered here, a Magnus expansion~\cite{MagnusReview} of the 
propagator is appropriate. We rely on the leading Magnus expansion term, which can be written as
\begin{equation}
    \hat{U}(T)\approx \exp[-i T H_\text{avg}]\ ,
\end{equation}
with 
\begin{equation}
H_\text{avg}=\frac{1}{T}\int_0^T H(t')dt'
\end{equation}
For the ramp shape considered here, $H_\text{avg}$ is independent of $T$ and is of the same
form as (\ref{eqn:simpleham}) with $U_{ij}$ and $\Omega$ unchanged, while $\delta(t)$ is replaced
by $\delta_\text{avg}=(\delta_0 + \delta_\text{final})/2$.

We now calculate symbolic expressions for the power series in $T$ of the connected correlators~(\ref{eqn:appg2}) 
relying on the propagator in the leading Magnus expansion 
form and starting from an initial state with all 
sites in the atomic ground state. When ($UT/\hbar,\Omega T,\delta T \ll 1$), 
we find the following expressions for the leading order in $T$ for a {\em single} 
path on a lattice linking sites $(0,0)$ and $(k,l)$ (for the sake of simplicity the results  
are only given for \mbox{$U_{ij}=U$}):
\begin{itemize}
\item{nearest neighbor correlation $|k|+|l|=1$}
    \begin{equation}
        g^{(2)}(T) = -\frac{1}{288} \left(U^2 - 
        3 U \delta_\text{avg}\right) \Omega^4 T^6,
    \end{equation}
\item{next nearest neighbor correlation $|k|+|l|=2$}
    \begin{align}
        g^{(2)}(T) &= \frac{ \Omega^6 T^{10} }{2419200} \left(77 U^4 - 340 U^3 \delta_\text{avg} + 
        375 U^2 \delta_\text{avg}^2\right),
    \end{align}
\item{third nearest neighbor correlation $|k|+|l|=3$}
    \begin{align}
        g^{(2)}(T) = -\frac{\Omega^8 T^{14}}{26824089600} \left(4279 U^6 - 24766 U^5 \delta_\text{avg} \right. \nonumber\\
            \left. + 46725 U^4 \delta_\text{avg}^2 - 28350 U^3\delta_\text{avg}^3 \right).
    \end{align}
\end{itemize}

\begin{figure}[t]
\includegraphics[width=8cm]{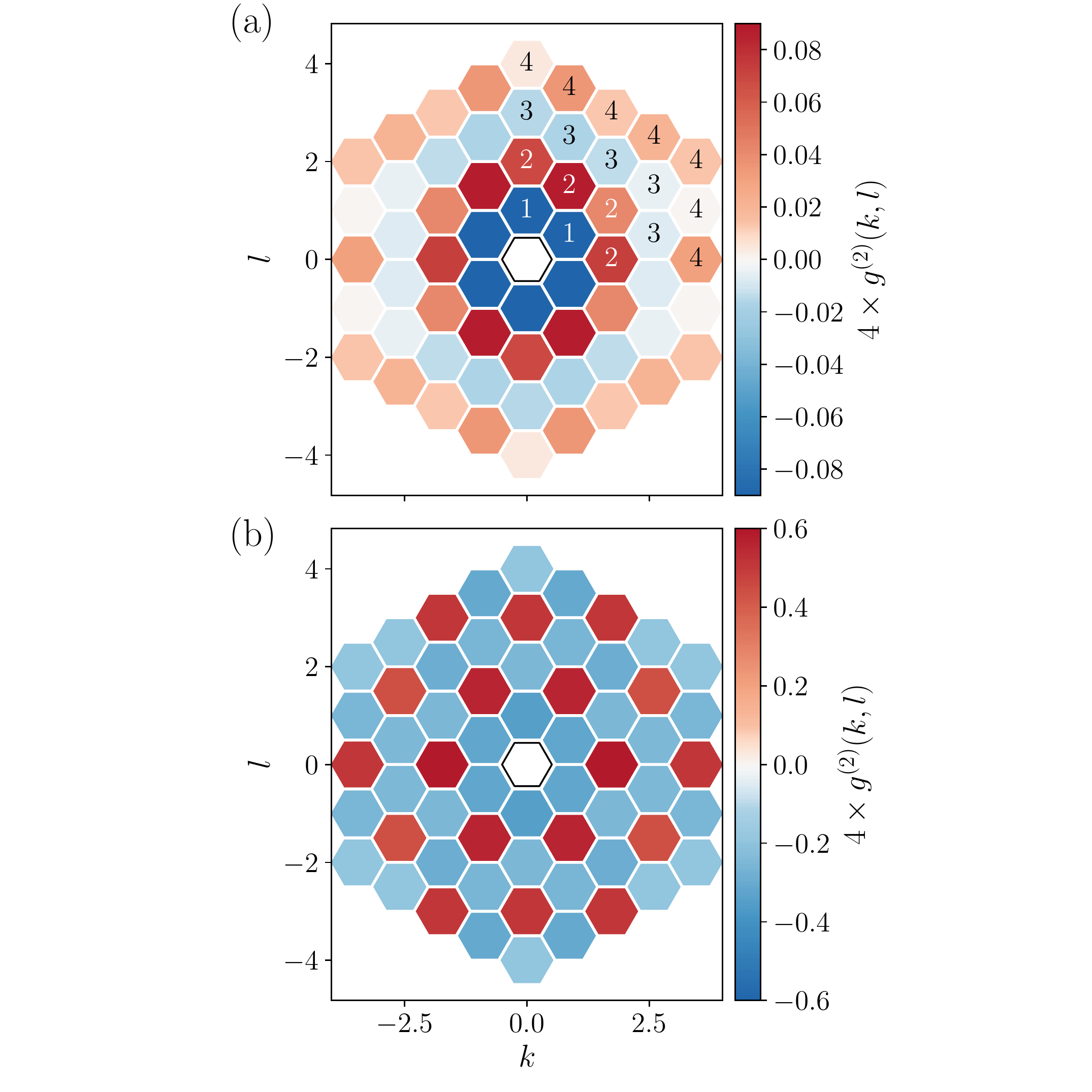}
\caption{Correlations on the $N=36$-site triangular array (only the first four Manhattan shells are shown). 
(a) Experimentally measured correlations after a sweep on the triangular
array featuring a pronounced Manhattan shell structure $g^{(2)}(k,l) \propto (-1)^m$.  
The numbers give the value of $m=|k|+|l|$. 
    (b) Correlations of the equal superposition of the classical groundstates
    for $\delta_{\text{final}}$ as reached at the end of the sweep used in (a).
    These correlations are in qualitative agreement with the correlations expected in 
the $1/3$ AF state in the right panel of Fig.~\ref{fig:fig1}(c).}
\label{fig:corrs_comp_tri}
\end{figure}

While the expressions look complicated, a common structure emerges: 
the leading $T$ dependence for a Manhattan distance $m=|k|+|l|$ is proportional to
${T^{2+4 m}}$. Neglecting the actual value of the coefficients for 
the moment in an admittedly crude approximation, these results suggest that
correlations at a next shell are suppressed by a factor $T^4$ 
compared to the previous shell, suggesting an exponential spatial decay of the
connected correlations at short times with $m$. 
These results also provide an insight as to why more distant shells require longer times to develop
appreciable correlations: they require larger values of $T$ to overcome the initial high powers in $T$ suppression of the correlations.

Importantly, the neglected higher-order terms in the Magnus expansions do not alter the leading order in $T$ for the considered ramp shapes, but only
affect the sub-leading $T$-coefficients. Finally, we note that at very short times, correlations induced by the direct van der Waals tail of the
interactions compete with the high-order nearest-neighbor considerations here. Although strictly speaking this leads to a breakdown of the exponential suppression of correlations beyond the light-cone, the smallness of this effect prevents its observation in the present experiments.

\subsection{Lattice embedding}
\label{app:short:time:embedding}

The expressions derived above are lattice independent, as long as there is a 
chain of $m$ ($m=|k|+|l|$ above) successive nearest neighbor interactions linking the
two considered sites~\cite{oitmaa_hamer_zheng_2006}. These expressions can now be embedded in {\em any} lattice (e.g.~cubic, kagome, honeycomb, ...), and the actual coefficients of the
short-time series can be derived by determining the corresponding embedding coefficients. The embedding counting is illustrated in Fig.~\ref{fig:binomialcoeffs} for
the square and the triangular lattice, yielding binomial coefficients multiplying the symbolic expressions for a single path derived above. Note that for the case $U_z \neq U_w$
the coefficients of the single paths can already differ before taking the embedding factors into account.

\section{Correlations on the triangular lattice: Short-time versus ground state correlations}
\label{app:corrstriangular}

The short-time expansion on the square lattice produces the same staggered correlation 
pattern as the one of the ground-state of an AF ordered phase. There is thus a similarity between the sign structure of the correlations at short times and at very long times (i.e.~the adiabatic limit).

On the triangular lattice, however, the  correlations obtained in the experiment [see \figref{fig:corrs_comp_tri}(a)], which  are in good agreement with the short-time expansion as shown in Fig.~\ref{fig:fig5}(c-d), differ significantly from the ones expected for the AF ground state with density $1/3$ [Fig.~\ref{fig:corrs_comp_tri}(b)]. The origin of this discrepancy may be related to the frustrated character of the triangular geometry and will be the subject of future work.


\begin{thebibliography}{62}
\bibitem [{foo()}]{footnote_nn_only}%
  \BibitemOpen
  \href@noop {} {}\bibinfo {note} {The next-nearest neighbor interaction
  strength is $(\sqrt{2})^6 = 8$ times smaller than the NN one and is thus one
  order of magnitude smaller than the Rabi frequency drive $\Omega$ and the
  inverse experimental time scale $1/t_\text{tot}$}\BibitemShut {NoStop}%
\end{thebibliography}

%

\end{document}